\documentclass[twocolumn,showpacs,aps,prl,superscriptaddress,groupedaddress]{revtex4-1}
\usepackage[english]{babel}
\usepackage[utf8]{inputenc}
\usepackage[T1]{fontenc}

\usepackage{hyperref}
\usepackage{amsmath, amssymb, amsfonts, mathrsfs}
\usepackage{xcolor}
\usepackage{exscale}
\usepackage{graphicx}
\usepackage{latexsym}

\usepackage{dsfont}

\newcommand{\ket}[1]{|#1\rangle}
\newcommand{\proj}[1]{|#1\rangle\langle#1|}
\DeclareMathOperator{\Tr}{Tr}
\DeclareMathOperator{\Id}{\mathds{1}}

\newcommand{\map}[1]{\mathscr{#1}}
\newcommand{\cnot}{\text{CNOT}}

\begin{document}

\title{Experimental Detection of Quantum Channels}

\author{Adeline Orieux$^1$}\email{adeline.orieux@gmail.com, group's webpage: http://quantumoptics.phys.uniroma1.it/}
\author{Linda Sansoni$^1$}
\author{Mauro Persechino$^1$}
\author{Paolo Mataloni$^{1,2}$}
\affiliation{$^1$Dipartimento di Fisica, Sapienza Università di Roma, Piazzale Aldo Moro, 5, I-00185 Roma, Italy\\
$^2$Istituto Nazionale di Ottica, Consiglio Nazionale delle Ricerche (INO-CNR), Largo Enrico Fermi, 6, I-50125 Firenze, Italy}

\author{Matteo Rossi$^3$}
\author{Chiara Macchiavello$^{3}$}
\affiliation{$^3$Dipartimento di Fisica and INFN-Sezione di Pavia, via Bassi 6, I-27100 Pavia, Italy}

\date{\today}

\begin{abstract}
We demonstrate experimentally the possibility of efficiently detecting properties of quantum channels and quantum gates. The optimal detection scheme is first achieved for non entanglement breaking channels of the depolarizing form and is based on the generation and detection of polarized entangled photons. We then demonstrate channel detection for non separable maps by considering the CNOT gate and employing two-photon hyperentangled states.
\end{abstract}

\maketitle

\textit{Introduction ---} The experimental realization of a quantum channel is unavoidably affected by noise. One possible way to check how well this has been performed is to make a full tomography of the process. This nevertheless is known to be very expensive in terms of number of measurements to be performed \cite{NC}. In many practical situations, however, one is only interested in some specific properties of the experimental channel, e.g. whether it has some entangling power, in order for the channel to be useful for a specific task, as e.g. quantum communication.

In this work we address this problem experimentally, following the method of quantum channel detection recently proposed in Refs. \cite{ns1,ns2}. The method allows us to detect properties of quantum channels when some \textit{a priori} information about the form of the channel is available. Besides being less informative than full process tomography, the method gives the advantage to single out the property of interest with a much smaller experimental effort than in the full tomography case. 

The method relies on the concept of witness operators \cite{horo-ter} and the Choi-Jamiolkowski isomorphism \cite{jam}. We briefly remind both of them. A state $\rho$ is entangled if and only if there exists a hermitian operator $W$ such that $\Tr[W\rho]< 0$ and $\Tr[W\rho_{sep}]\geq 0$ for all separable states; such an operator is called an entanglement witness. The Choi-Jamiolkowski isomorphism gives a one-to-one correspondence between completely positive (CP) and trace-preserving (TP) maps acting on $\mathcal{D(H)}$ (the set of density operators on $\mathcal{H}$, with finite dimension $d$) and bipartite density operators $C_{\map{M}}$ on $\mathcal{H\otimes H}$ (named Choi states). The isomorphism can be stated as 
\begin{equation}
\map{M}\Longleftrightarrow C_{\map{M}}=(\map{M}\otimes\map{I})[\proj{\alpha}],
\end{equation}
where $\map{I}$ is the identity map, and $\ket{\alpha}$ is the maximally entangled state with respect to the bipartite space $\mathcal{H\otimes H}$, i.e. $\ket{\alpha}=\frac{1}{\sqrt d}\sum_{k=1}^d\ket{k}\ket{k}$. The above isomorphism can be exploited to link convex sets of quantum channels to particular sets of quantum states. In the following the proposed method will be applied to the convex sets of either entanglement breaking (EB) channels and separable channels.

\textit{1-qubit EB channels: Theory ---} A channel $\map{M}$ is EB if and only if its Choi state $C_\map{M}$ is separable \cite{EB}. Therefore, the detection of entanglement of $C_\map{M}$ in the doubled system by using a witness operator $W_{EB}$ suitable for $C_{\map{M}}$ allows us to prove that the implemented quantum channel $\map{M}$ was not EB.

We will show the method for the depolarizing channel acting on one qubit, defined as
\begin{equation}\label{depo}
\Gamma_p[\rho]= \sum^3_{i=0}{p_i \sigma_i \rho \sigma_i},
\end{equation}
where $\sigma_0$ is the identity operator, $\{\sigma_i\}_{i=1,2,3} $ are the three Pauli operators $\sigma_x,\sigma_y, \sigma_z$ respectively, and $p_0=1-p$ (with $p\in[0,1]$), while $p_i=p/3$ for $i=1,2,3$.
Such a channel is known to be EB only for $p \geq 1/2$. Denoting the maximally entangled state of two qubits as $\ket{\Phi^+}$, the corresponding Choi state is given by
\begin{equation}\label{werner states}
C_{\Gamma_p}= (1-\frac{4}{3}p)\proj{\Phi^+}+\frac{p}{3}\Id\;,
\end{equation}
which leads \cite{ent-wit,jmo} to a suitable detection operator of the form \cite{ent-wit,ns2}
\begin{equation}\label{wEB}
W_{EB}= \frac{1}{4}(\Id\otimes\Id -\sigma_x \otimes \sigma_x + \sigma_y \otimes \sigma_y -\sigma_z \otimes \sigma_z)\;.
\end{equation}

\begin{figure}[h]
\centering
\includegraphics{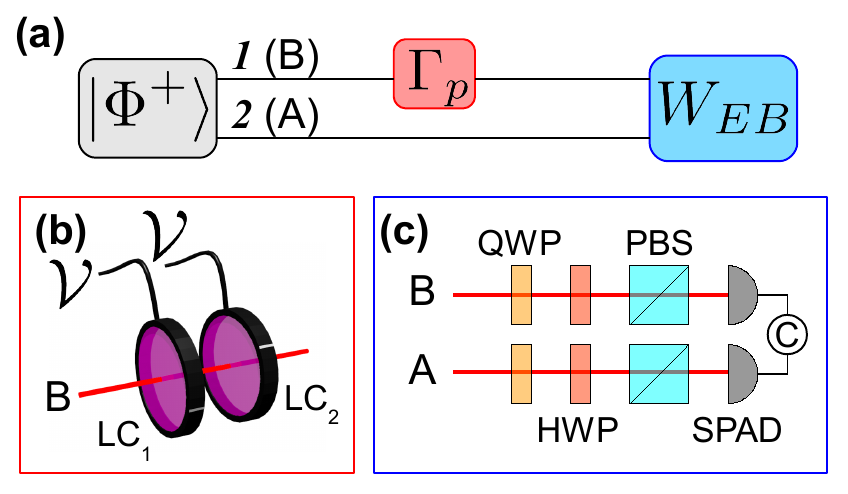}
\caption{(color online) (a) Scheme for the 1-qubit depolarizing channel detection; $\ket{\Phi^+}$: 2-qubit entangled state, $\Gamma_p$: 1-qubit depolarizing channel; $W_{EB}$: EB witness measurement. (b) Implementation of the 1-qubit depolarizing channel; LC$_{1,2}$: liquid crystal retarders with axis set at 0$^{\circ}$ and 45$^{\circ}$ respectively; $\mathcal{V}$: applied voltage to the LCs. (c) Polarization analysis set-up used to evaluate the witness $W_{EB}$; QWP: quarter-wave plate, HWP: half-wave plate, PBS: polarizing beam-splitter, SPAD: single-photon avalanche photodiode, C: coincidence counting electronics.}
\label{FigPauli}
\end{figure}

The detection scheme is depicted in Fig. \ref{FigPauli} (a): we prepare the two-qubit system in the maximally entangled state $\ket{\Phi^+}$, we then let the depolarizing channel act on qubit \textit{1}, and we finally measure the operator $W_{EB}$ acting on both qubits at the end. If $\langle W_{EB}\rangle<0$, then we are guaranteed that the depolarizing channel $\Gamma_p$ is not EB. Notice that the theoretical calculated expectation value for the ideal Choi state is $\langle W_{EB}\rangle=p-1/2$, that guarantees the detection of all non EB depolarizing channels because it gives a negative expectation value whenever $p<1/2$.

Notice that from the measured $\langle W_{EB}\rangle$ we can establish a lower bound \cite{ns2} on the theoretical quantity $\mu_c(\Gamma_p)$ introduced in \cite{giova}, which represents the minimal amount of noise we need to add to $\Gamma_p$ via a classical stochastic process  in order to make the resulting map EB.
Such a bound is given by
\begin{equation}
\mu_c(\Gamma_p)\geq\frac{2|\langle W_{EB}\rangle|}{1+2|\langle W_{EB}\rangle|}.
\end{equation}

\textit{1-qubit EB Channels: Experiment ---} The two-photon states used in this work were produced by a spontaneous parametric down conversion (SPDC) source operating on the double excitation (back and forth) of a type I, $0.5\,mm$-long BBO crystal, that, depending on the performed experiment, allows to generate either a polarization entangled state \cite{PRA70rome}, or a path-polarization hyperentangled state \cite{PRA72rome} of two photons emitted over either two or four spatial modes (see Supplementary Information (SI) for major details).

For this experiment, the 2-photon polarization entangled state generated over two spatial modes (Fig. \ref{FigPauli} (a)) was: $|\Phi^+\rangle=\frac{1}{\sqrt{2}}\left(|H\rangle_B|H\rangle_A+|V\rangle_B|V\rangle_A\right)$, where $H$ ($V$) stands for the horizontal (vertical) polarization of photon $A$ (Alice's) or $B$ (Bob's).

We simulated a 1-qubit depolarizing channel, Eq. \eqref{depo}, acting on Bob's photon by inserting two liquid crystal retarders (LC$_{1}$ and LC$_{2}$) on the path of photon $B$, one having its fast axis horizontal and the other oriented at 45$^{\circ}$ with respect to the horizontal \cite{PRL107romepavia} (Fig. \ref{FigPauli} (b)). Depending on the applied voltage $\mathcal{V}$, it is possible to change the retardation between ordinary and extraordinary polarized radiation. More precisely, by applying either $\mathcal{V}_ {\mathds{1}}$ or $\mathcal{V}_{\pi}$ to a LC, it can be made to act as either a full- or a half-wave plate, respectively. Thus, by varying independently the voltage applied to LC$_{1}$ and LC$_{2}$ for different time intervals, we could apply the four Pauli operators to photon $B$ with different values of the weight $p$ (see SI).

To measure the witness $W_{EB}$ given by Eq. \eqref{wEB} as a function of the noise level, varying between the values 0 and 1, we needed to evaluate $\sigma_x^{\otimes 2}$, $\sigma_y^{\otimes 2}$ and $\sigma_z^{\otimes 2}$ for different values of $p$. This was done, for each choice of $p$, by measuring the coincidences between photons $A$ and $B$ in 8 different settings \cite{PRL91romepavia} of the polarization analysis set-up which consisted of a quarter-wave plate (QWP), a half-wave plate (HWP), a polarizing beam-splitter (PBS) and a single-photon avalanche photodiode (SPAD) (Fig. \ref{FigPauli} (c)).

\begin{figure}[h]
\centering
\includegraphics{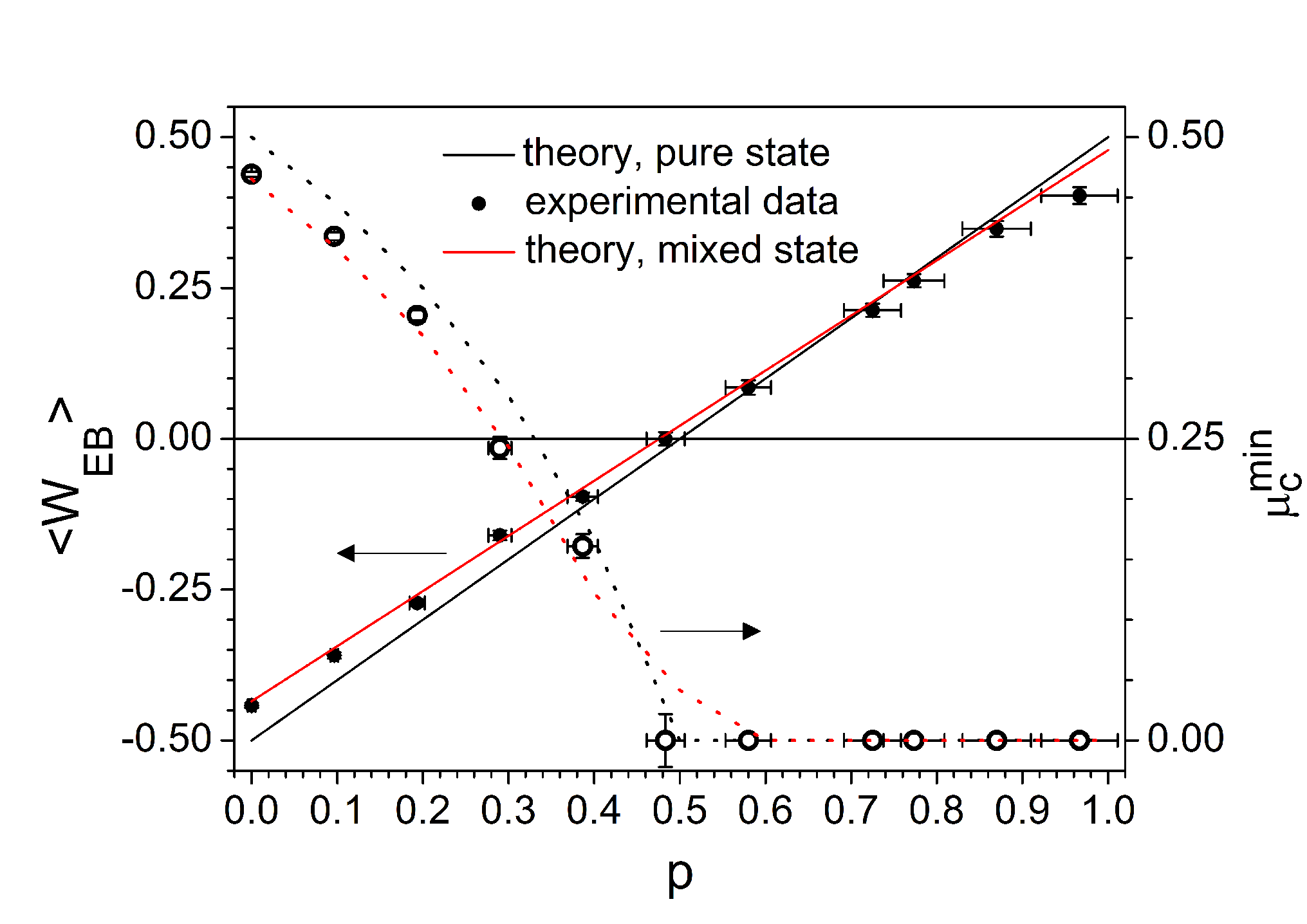}
\caption{(color online) Witness value $\langle W_{EB}\rangle$ (full symbol, solid line) and minimal bound of $\mu_c(\Gamma_p)$ (empty symbol, dotted line) as a function of the noise parameter $p$.}
\label{Fig2}
\end{figure}

The witness we obtained is shown in Fig. \ref{Fig2}, together with the theoretical behaviour for a perfectly pure state and the actual one used in the experiment. To compare our results with the theory, we need in fact to take into account the imperfection of the experimentally simulated Choi state. Indeed, the two-photon state produced by the SPDC source corresponds to $|\Phi^+\rangle$ only up to a finite fidelity $F_0=0.935\pm0.004$ (measured by performing a two-photon quantum state tomography for $p=0$). Replacing $|\Phi^+\rangle\langle\Phi^+|$ by $F_0\,|\Phi^+\rangle\langle\Phi^+|+\frac{1-F_0}{3}(|\Phi^-\rangle\langle\Phi^-|+|\Psi^+\rangle\langle\Psi^+|+|\Psi^-\rangle\langle\Psi^-|)$ in Eq. \eqref{werner states}, we can thus write the experimental Choi state as:
\begin{eqnarray}
C_{\Gamma_p,exp} &=& \Big(1-\frac{4\,p}{3}\Big)\frac{4\,F_0-1}{3}|\Phi^+\rangle\langle\Phi^+| \nonumber\\
&+& \Big(\frac{p}{3}\frac{4\,F_0-1}{3}+\frac{1-F_0}{3}\Big)\Id.
\end{eqnarray}
The error bars on $\langle W_{EB}\rangle$ are obtained by propagating the Poissonian uncertainties associated with the coincidence counts and the error bars on $p$ are estimated by considering the finite response time of the LC.

Let us notice that we indeed obtain the entanglement breaking property of the channel for a value of $p$ up to around 0.5 as expected from the theory, and as a consequence the bound on $\mu_c(\Gamma_p)$ gets trivial above this value (see Fig. \ref{Fig2}).

\textit{2-qubit separable maps: Theory ---} We will now consider the set of separable maps acting on bipartite systems, that are defined as 
\begin{equation}\label{SRU}
\map{S}[\rho_{\textit{12}}]=\sum_k (A_k\otimes B_k)\rho_{\textit{12}} (A_k\otimes B_k)^\dagger,
\end{equation}
where $A_k$ and $B_k$ act on systems \textit{1} and \textit{2} respectively. As for EB channels, the set of separable maps $\map{S}$ is convex, and it is then possible to detect a general map lying outside it. Notice that, since the regarded maps now act on a bipartite state $\rho_{\textit{12}}$, the corresponding Choi states refer to a four-partite system \textit{1234} which is separable in the splitting \textit{13}|\textit{24} \cite{sep1,kraus_sep}. As a demonstration of the achievability of the optimal detection method for non separable maps we will consider the explicit case of the CNOT gate. The corresponding detection operator is given by \cite{ns1,ns2}
\begin{equation}\label{WCNOT}
W_\cnot=\frac{1}{2}\Id-\proj{\cnot},
\end{equation}
where $\ket{\cnot}$ is the Choi state associated to the gate CNOT (with qubit \textit{1} as the target and qubit \textit{2} as the control, and $\ket{\alpha}=\ket{\Phi^+}_{\textit{13}}\ket{\Psi^+}_{\textit{24}}$ \cite{note}), namely
\begin{equation}\label{cnotstate}
\ket{\cnot}=\frac{1}{\sqrt 2}(\ket{\Phi^+}_{\textit{13}}\ket{01}_{\textit{24}}+\ket{\Psi^+}_{\textit{13}}\ket{10}_{\textit{24}})\;,
\end{equation}
where $\ket{\Phi^+}$ and $\ket{\Psi^+}$ are maximally entangled states of the Bell basis.
The witness above can be measured by using nine different local measurement settings \cite{ns1,ns2}. A possible way to reduce the experimental effort is to consider the suboptimal operator \cite{ns2} 
\begin{eqnarray}\label{WCNOTsub}
\tilde W_\cnot=3\Id &-&2\left[\frac{(\Id+\Id \sigma_x^{\otimes 3})}{2} \frac{(\Id+ \sigma_x\Id \sigma_x\Id)}{2}\right.\nonumber\\
&+&\left.\frac{(\Id- \Id \sigma_z\Id \sigma_z)}{2}\frac{(\Id +\sigma_z^{\otimes 3}\Id)}{2}\right],
\end{eqnarray}
where we omitted the tensor products and from which it is clear that it requires only two measurement settings.

\begin{figure}[h]
\centering
\includegraphics{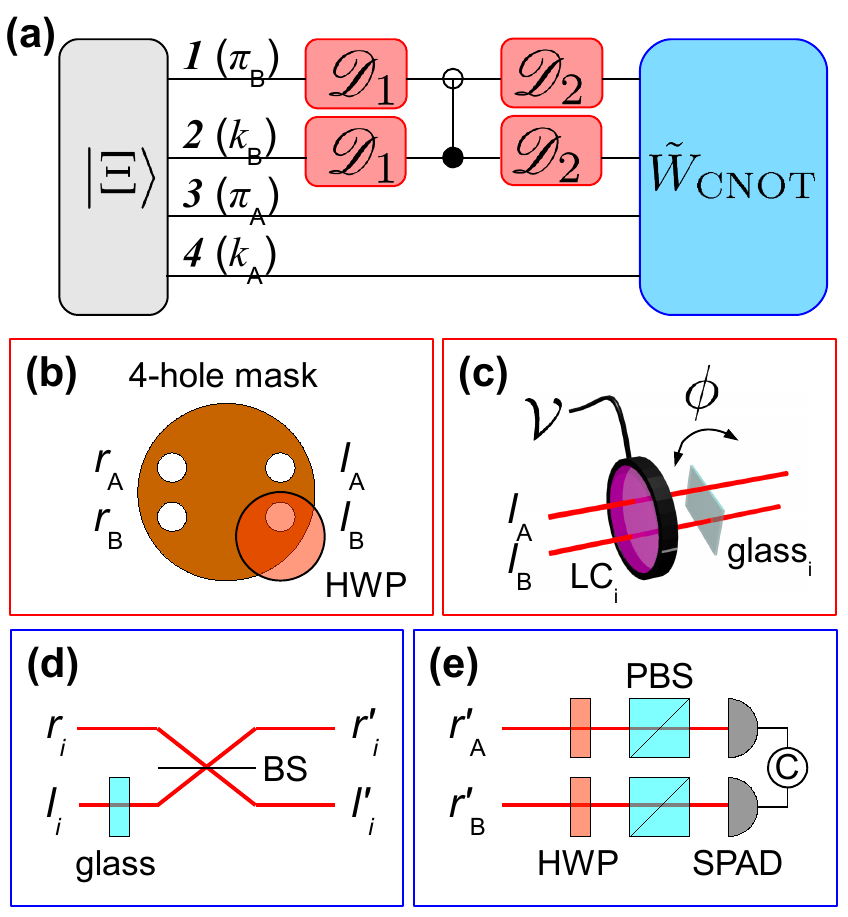}
\caption{(color online) (a) Scheme for the 2-qubit CNOT channel detection in presence of dephasing noise; $|\Xi\rangle$: 4-qubit hyperentangled state; $\pi_{i}$: polarization qubit and $k_{i}$: path qubit, with $i=A,B$; $\map{D}_{1,2}$: independent 2-qubit dephasing channels; $\tilde W_\cnot$: CNOT witness measurement. (b) CNOT implementation: a half-wave plate (HWP) set at 45$^{\circ}$ flips the polarization of photon $B$ when its path is $l_B$. (c) 2-qubit dephasing channel implementation; LC$_{i}$: liquid crystal retarder with its axis set at 0$^{\circ}$; glass$_i$: thin glass plate; $i=1,2$. (d) Path analysis set-up; glass: thin glass plate, BS: beam-splitter; $i=A,B$. (e) Polarization analysis set-up used in combination with (c) to evaluate the witness $\tilde W_\cnot$; HWP: half-wave plate, PBS: polarizing beam-splitter, SPAD: single-photon avalanche photodiode, C: coincidence counting electronics.}
\label{FigCNOT}
\end{figure}

In this work we also demonstrate the robustness of the detection method in the presence of dephasing noise, which is of the form \eqref{depo} with $p_0=1-q_i$, $p_1=p_2=0$ and $p_3=q_i$, $i=1,2$. We consider the case where the dephasing noise acts on both qubits, before and/or after the CNOT gate, as follows (Fig. \ref{FigCNOT} (a)):
\begin{equation}\label{depolarising}
\map{M}_{\cnot,\map{D}}=(\map{D}_2\otimes\map{D}_2)\cnot(\map{D}_1\otimes\map{D}_1)
\end{equation}
Notice that the four dephasing processes act independently and are assumed to have the same strength ($q_1$ ($q_2$) before (after) the CNOT gate) for the two qubits.

The noise robustness of the operator $\tilde W_\cnot$ with respect to dephasing noise is evaluated by the expectation value of $\tilde W_\cnot$ given by Eq. \eqref{WCNOTsub} with respect to the state $C_{\map{M}_{\cnot,\map{D}}}$ (the Choi state corresponding to the composite map $\map{M}_{\cnot,\map{D}}$).
We stress that, despite it requires only two measurement settings, the witness $\tilde W_\cnot$ of Eq. \eqref{WCNOTsub} turns out to be as efficient as $W_\cnot$ of Eq. \eqref{WCNOT} in the presence of dephasing noise, since the two operators detect non separability of $\map{M}_{\cnot,\map{D}}$ in the same range of values of the noise parameters. Therefore, in the present experiment we measure $\tilde W_\cnot$ instead of $W_\cnot$. The theoretical expectation value is given by \cite{ns2}:
\begin{eqnarray}\label{thres_dephase}
\Tr[\tilde W_\cnot C_{\map{M}_{\cnot,\map{D}}}] &=& 1-2[(1-q_1)^2(1-q_2)^2 \nonumber\\
&+& q_1q_2(1-q_1q_2)].
\end{eqnarray}
The roots of the above expression define the threshold values for the noise parameters in order to have a successful non separability detection for the noisy map $\map{M}_{\cnot,\map{D}}$. In case the noise has the same strength before and after the CNOT gate ($q_1=q_2=q$), it is possible to detect the non separability character of the map for sufficiently low values of the noise parameter: $q<0.17$. (The case $q_1 \neq q_2$ is further studied in the SI.)

\textit{2-qubit separable maps: Experiment ---} For this second experiment, we used the SPDC source operating over four emission modes (see SI). Hence we prepared the 4-qubit hyperentangled state $|\Xi\rangle=|\Phi^+\rangle_{\textit{13}}|\Psi^+\rangle_{\textit{24}}$ where $|\Phi^+\rangle_{\textit{13}}=\frac{1}{\sqrt{2}}\left(|H\rangle_{B}|H\rangle_{A}+|V\rangle_{B}|V\rangle_{A}\right)$ and $|\Psi^+\rangle_{\textit{24}}=\frac{1}{\sqrt{2}}\left(|r\rangle_{B}|l\rangle_{A}+|l\rangle_{B}|r\rangle_{A}\right)$, where $r$ ($l$) designs the right (left) path of photon $A$ or $B$.

We implemented a CNOT gate on Bob's photon by inserting a half waveplate set at 45$^{\circ}$ on the left path of photon $B$: thus the path (qubit \textit{2}) acts as the control and the polarization (qubit \textit{1}) acts as the target (Fig. \ref{FigCNOT} (b)). After the CNOT gate, the 4-qubit state then reads:
\begin{equation}\label{Xi_out}
\begin{split}
|\Xi_{out}\rangle_{\textit{1234}}&=\frac{1}{2}\big(|H\,r\rangle_{B}|H\,l\rangle_{A}+|V\,l\rangle_{B}|H\,r\rangle_{A}\\&\quad+|V\,r\rangle_{B}|V\,l\rangle_{A}+|H\,l\rangle_{B}|V\,r\rangle_{A}\big)\\&=\frac{1}{\sqrt{2}}\left(|\Phi^+\rangle_{\textit{13}}|r\rangle_\textit{2}|l\rangle_\textit{4}+|\Psi^+\rangle_{\textit{13}}|l\rangle_\textit{2}|r\rangle_\textit{4}\right).
\end{split}
\end{equation}
Using the correspondence $|H\rangle_{B,A}\leftrightarrow |0\rangle_{\textit{13}}$,$|V\rangle_{B,A}\leftrightarrow |1\rangle_{\textit{13}}$ , $|r\rangle_{B,A}\leftrightarrow |0\rangle_{\textit{24}}$ and $|l\rangle_{B,A}\leftrightarrow |1\rangle_{\textit{24}}$, Eq. \eqref{Xi_out} is equivalent to the Choi state of the CNOT channel expressed in the logical basis \eqref{cnotstate}.

Dephasing noisy channels were simulated by acting independently on qubits \textit{1} and \textit{2}, before and/or after the CNOT gate, as in Eq. \eqref{depolarising}, by inserting a LC with its fast axis at 0$^{\circ}$ with respect to the horizontal and a thin glass plate, both before and after the CNOT (Fig. \ref{FigCNOT} (c)). Each LC induces a phase between $|H\rangle_B$ and $|V\rangle_B$, that can be set to either 0 or $\pi$ by applying a voltage $\mathcal{V}_ {\mathds{1}}$ or $\mathcal{V}_ {\pi}$ respectively, thus acting either as $\mathds{1}$ or $\sigma_z$ for qubit \textit{1}; each glass plate introduces a phase $\phi$ between $|r\rangle_B$ and $|l\rangle_B$, that can be set to 0 or $\pi$ by calibrated rotations of the plate, thus acting either as $\mathds{1}$ or $\sigma_z$ for qubit \textit{2}. By varying the relative time of action of each dephaser, in a similar manner as in the 1-qubit channel experiment, we were able to vary the values of $q_1$ and $q_2$. 

To measure the witness $\tilde W_{\cnot}$ \eqref{WCNOTsub} as a function of $q_1$ and $q_2$, we needed to evaluate $\sigma_x^{\otimes 4}$ and $\sigma_z^{\otimes 4}$ for several values of $q_1$ and $q_2$.
Thus, for each value of $q_1$ and $q_2$, we measured coincidence counts between photons $A$ and $B$ in 32 different settings of the polarization-path analysis set-up. The polarization analysis in this case is achieved via a HWP and a PBS (Fig. \ref{FigCNOT} (e)) while the path analysis is done either directly sending the photons to the detectors (thus measuring $|r\rangle$ and $|l\rangle$) or passing them first through a beam-splitter and a thin glass plate (thus measuring $|d\rangle=\frac{1}{\sqrt{2}}(|r\rangle+|l\rangle)$ and $|a\rangle=\frac{1}{\sqrt{2}}(|r\rangle-|l\rangle)$) (Fig. \ref{FigCNOT} (d)).

\begin{figure}[t]
\centering
\includegraphics{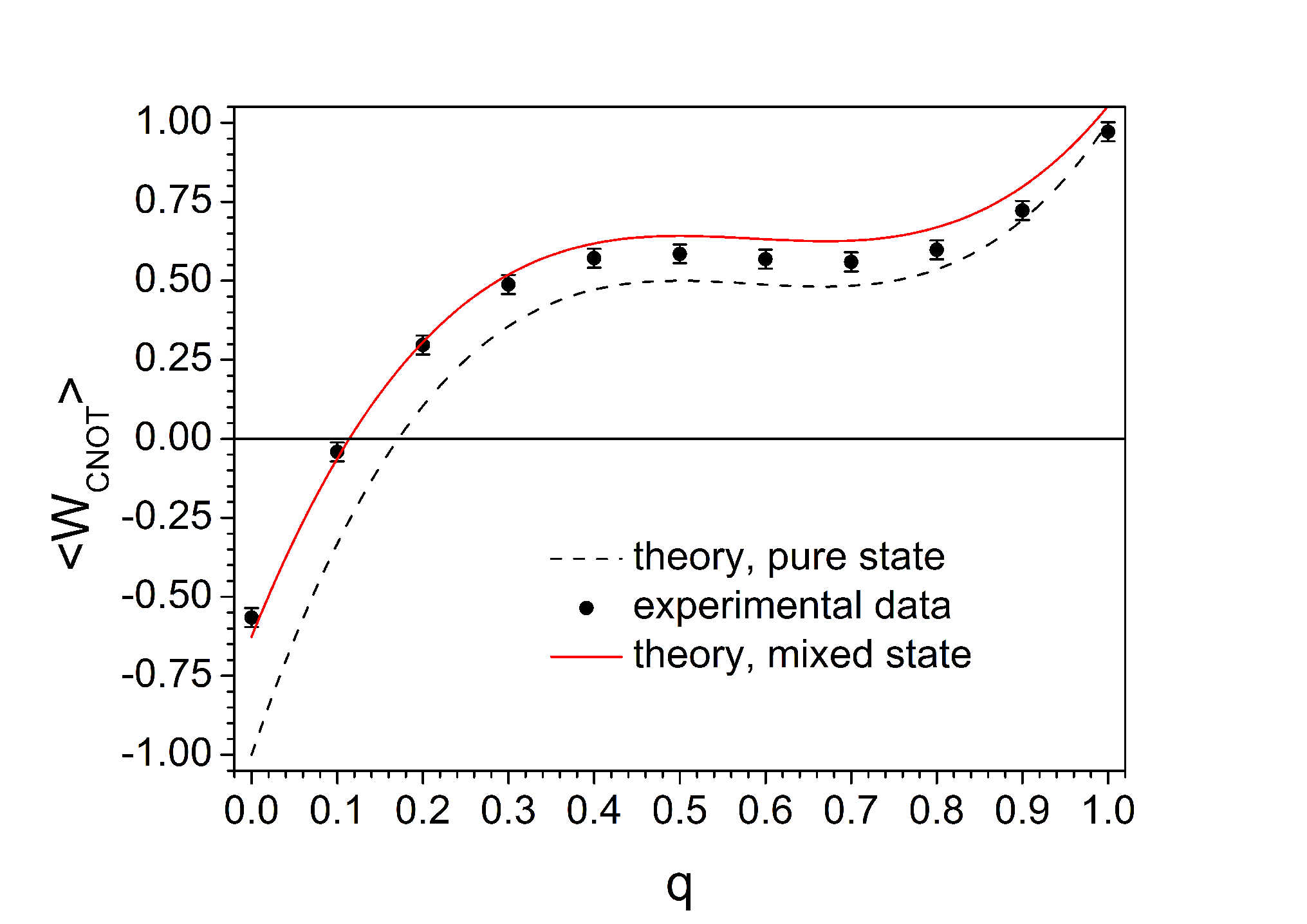}
\caption{(color online) CNOT witness value $\langle\tilde W_\cnot\rangle$ as a function of the noise parameter $q$.}
\label{Fig4}
\end{figure}

We obtained the witness values reported in Fig. \ref{Fig4} as a function of $q_1=q_2=q$. Again, to compare them properly with the theory, we must take into account the finite purity of the initial Choi state that we prepared to simulate the Choi state of the CNOT gate. We could model the experimental Choi state of the CNOT noisy channel, given the visibilities (measured in the diagonal basis) of the polarization ($\nu_{\pi}=0.858\pm 0.008$) and path ($\nu_k=0.904\pm 0.004$) entanglement for $q=0$ (see SI). As can be seen, our results are in good agreement with the theoretical calculation. Note that the slight discrepancy remaining for large $q$ is probably due to imperfections in the simulated dephasing channels. As expected, from these results it is evident that a low level of noise makes the CNOT to be no more an entangling gate, in particular the non separability of the map is no longer detected for $q > 0.1$ in our experiment.

\vspace{0.5 cm}
\textit{Conclusion ---} We have implemented a method that allows to check the entanglement properties of a noisy multi-qubit gate with fewer measurements than those required by a full quantum process tomography \cite{JOSAB24gatemeasure} and could thus be a more convenient tool for routine performance checks on quantum gates. This method has been tested in the cases of a 1-qubit entanglement breaking channel and of a 2-qubit separable map with very good agreement between experimental measurements and theoretical predictions.

The work was supported by the EU Project QWAD (Quantum Waveguides Applications \& Developments).

\end{document}


\title{Experimental Detection of Quantum Channels}

\author{Adeline Orieux$^1$}
\author{Linda Sansoni$^1$}
\author{Mauro Persechino$^1$}
\author{Paolo Mataloni$^{1,2}$}
\affiliation{$^1$Dipartimento di Fisica, Sapienza Università di Roma, Piazzale Aldo Moro, 5, I-00185 Roma, Italy\\
$^2$Istituto Nazionale di Ottica, Consiglio Nazionale delle Ricerche (INO-CNR), Largo Enrico Fermi, 6, I-50125 Firenze, Italy}

\author{Matteo Rossi$^3$}
\author{Chiara Macchiavello$^{3}$}
\affiliation{$^3$Dipartimento di Fisica and INFN-Sezione di Pavia, via Bassi 6, I-27100 Pavia, Italy}

\date{\today}

\maketitle

\section{Supplementary Information}

\subsection{Experimental setup}

\begin{figure}[h]
\centering
\includegraphics{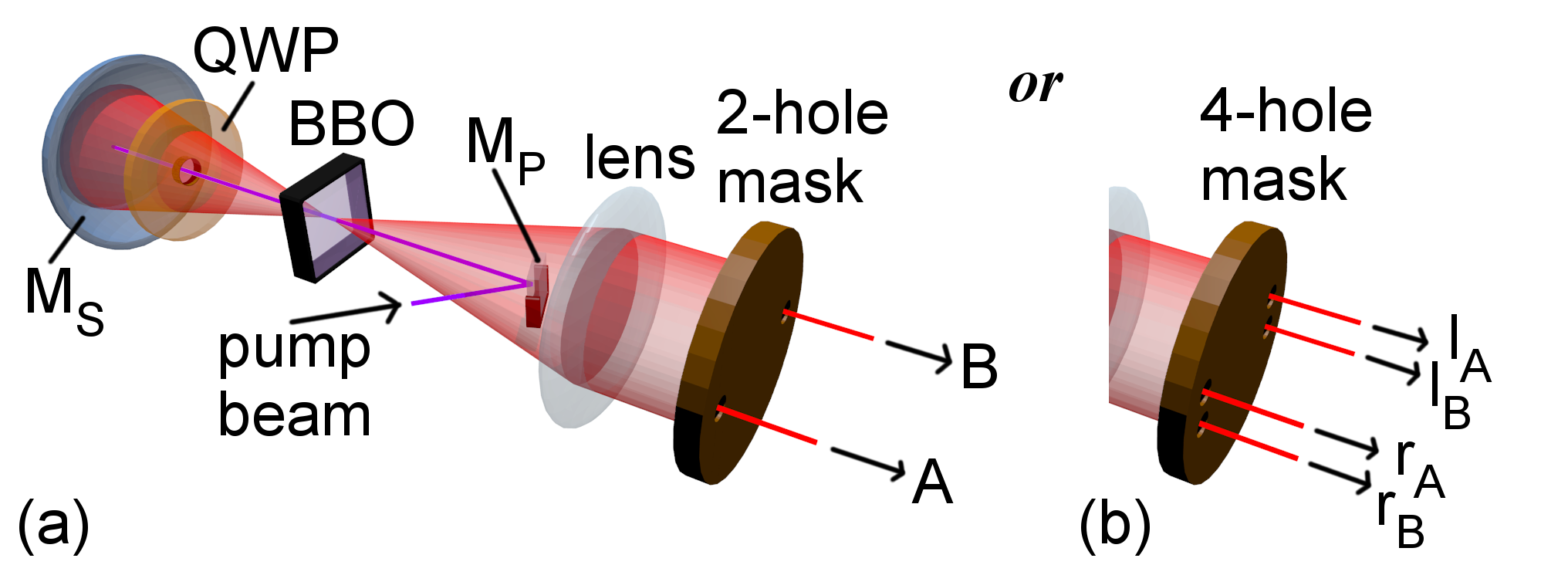}
\caption{(color online) Sketch of the SPDC source of two-photon polarization entangled states (a) or path-polarization hyper-entangled states (b). BBO: $\beta$-barium borate crystal, M$_{P}$: pump mirror, M$_{S}$: spherical mirror, QWP: quarter-wave plate.}
\label{Fig1}
\end{figure}

\textit{SPDC source ---}
The two-photon source used in this work is depicted in Fig. \ref{Fig1}: a type I, $0.5\,mm$-long BBO crystal, illuminated back and forth by a CW laser at $355\,nm$ generates two H-polarized cones of photon pairs by spontaneous parametric down-conversion (SPDC). One of the cones is transformed into a V-polarized cone by the combined action of a spherical mirror and a quarter-wave plate, thus allowing the production of polarization-entangled photon pairs, on two spatial modes selected by a 2-hole mask (Fig. \ref{Fig1} (a), \cite{PRA70rome}). Four spatial modes of emitted pairs ($r_A$-$l_B$ and $l_A$-$r_B$) can be selected by use of a 4-hole mask instead, thus allowing the generation of path-polarization hyper-entangled photon pairs (Fig. \ref{Fig1} (b), \cite{PRA72rome}).

\begin{figure}[h]
\centering
\includegraphics{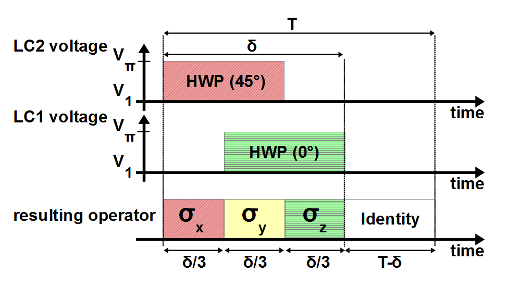}
\caption{(color online) Implemented Pauli operators during a detection gate. $\delta$: time when $\mathcal{V}_ {\pi}$ is applied to LC$_{1}$ and/or LC$_{2}$, $T$: detection gate duration; during $T-\delta$, $\mathcal{V}_ {\mathds{1}}$ is applied to both LC$_{1}$ and LC$_{2}$.}
\label{Fig2}
\end{figure}

\textit{Depolarizing channel ---}
The 1-qubit depolarizing channel acting on Bob's photon consists of two liquid crystal retarders (LC$_{1}$ and LC$_{2}$), the fast axis of LC$_{1}$ being horizontal and the one of LC$_{2}$ being oriented at 45$^{\circ}$. Thus, LC$_{1}$ acts as $\mathds{1}$ or $\sigma_z$ when the applied voltage is $\mathcal{V}_ {\mathds{1}}$ or $\mathcal{V}_ {\pi}$ respectively, LC$_{2}$ acts as $\mathds{1}$ or $\sigma_x$ when the applied voltage is $\mathcal{V}_ {\mathds{1}}$ or $\mathcal{V}_ {\pi}$ respectively, and they act as $\sigma_y$ when $\mathcal{V}_ {\pi}$ is applied to both. The weight $p$ is varied by changing the duration $\delta$ of the voltage $\mathcal{V}_ {\pi}$ on each LC with respect to the detection gate duration $T$: $p=\frac{\delta}{T}$ (Fig. \ref{Fig2}, \cite{PRL107romepavia}).

\begin{figure*}[t]
\centering
\includegraphics{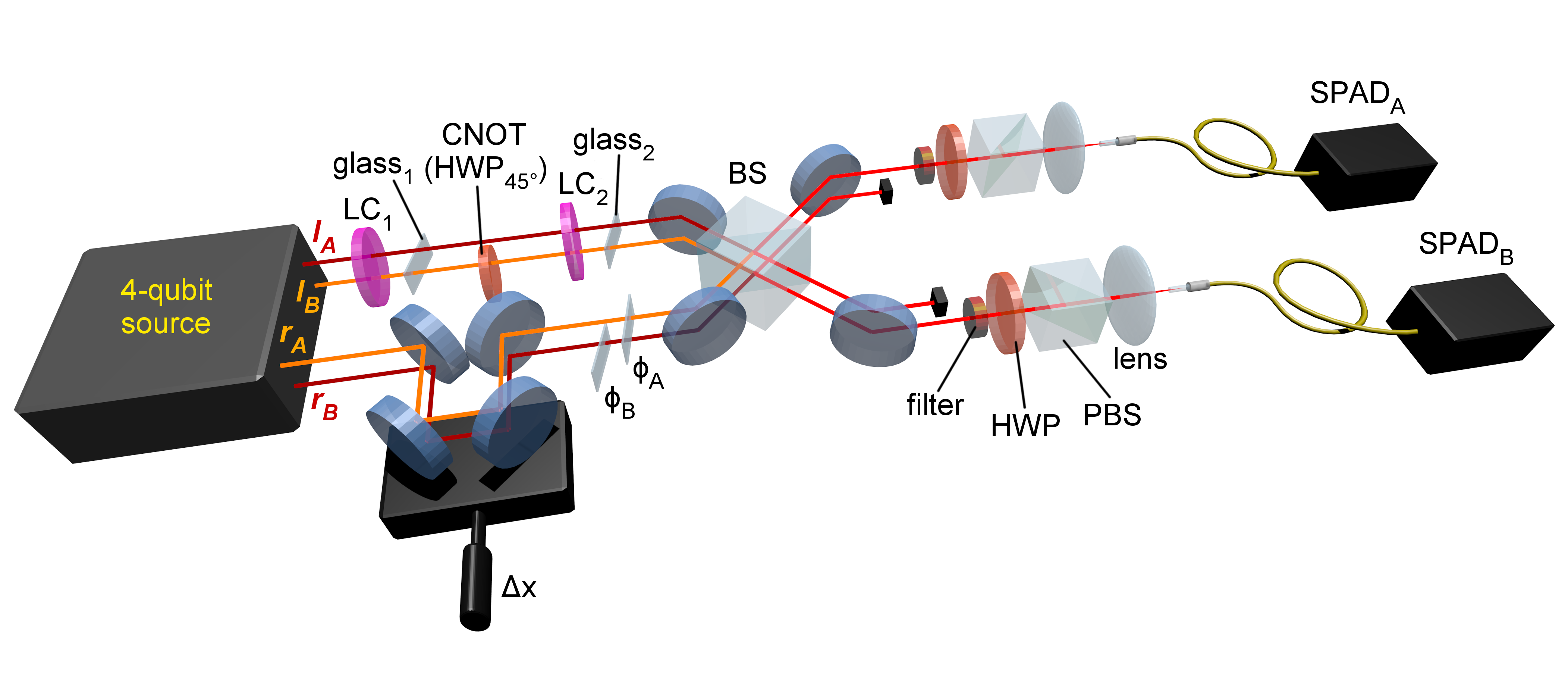}
\caption{(color online) Experimental set-up of the 2-qubit noisy CNOT channel. 4-qubit source: polarization-path hyperentangled two-photon source (Fig. \ref{Fig1} (b)); CNOT: half-wave plate set at 45$^{\circ}$; LC$_{1,2}$: liquid crystal retarders set at 0$^{\circ}$; glass$_{1,2}$: thin glass plates; BS: 50-50 beam-splitter; $\phi_A,B$: thin glass plates; $\Delta x$: adjustable delay between the left and the right path; HWP: half-wave plate; PBS: polarizing beam-splitter; SPAD: fibered single-photon avalanche photodiodes.}
\label{Fig3}
\end{figure*}

\textit{2-qubit separable map set-up ---}
The detailed experimental set-up of the 2-qubit separable map detection is shown in Fig. \ref{Fig3}. LC$_{1}$ (LC$_{2}$) and glass$_{1}$ (glass$_{2}$) implement a 2-qubit dephasing channel on photon $B$ before (after) the CNOT gate; the path qubits are analysed with a beam-splitter (BS) and two glass plates ($\phi_A$ and $\phi_B$) with a delay $\Delta x=0$ between the left and the right path; a half-wave plate (HWP) and a polarizing beam-splitter (PBS) are used to analyse the polarization qubits; the photons are detected in coincidence by two fibered single-photon avalanche photodiodes (SPAD).

Note that the Pauli operators for the polarization qubits \textit{1,3} may be expressed as: $\sigma_x^{(i)} = |D\rangle_i \langle D|_i- |A\rangle_i \langle A|_i$, $\sigma_y^{(i)} = |L\rangle_i \langle L|_i- |R\rangle_i \langle R|_i$, $\sigma_z^{(i)} = |H\rangle_i \langle H|_i- |V\rangle_i \langle V|_i$, where $D$, $A$, $L$ and $R$ design the linear diagonal, linear antidiagonal, circular left and circular right polarization, respectively, and $i=A,B$. For the path qubits \textit{2,4}, the Pauli operators may be written as: $\sigma_x^{(i)} = |d\rangle_i \langle d|_i- |a\rangle_i \langle a|_i$ and $\sigma_z^{(i)} = |r\rangle_i \langle r|_i- |l\rangle_i \langle l|_i$, where $|d\rangle_i=\frac{1}{\sqrt{2}}\left(|r\rangle_i+|l\rangle_i\right)$ and $|a\rangle_i=\frac{1}{\sqrt{2}}\left(|r\rangle_i-|l\rangle_i\right)$ are the diagonal and antidiagonal path states, and $i=A,B$.

\subsection{2-qubit separable map witness results}

\textit{Experimental CNOT Choi state model ---}
In order to model the imperfection of the experimental CNOT Choi state, we mainly consider two decoherence sources. Firstly, there is a depolarising process on the polarization degree of freedom given by the non perfect emission of the state $\ket{\Phi^+}_{\textit{13}}$ from the source. The noisy initial state can then be modelled as
\begin{equation}
\rho_{in}=(\nu_\pi\ket{\Phi^+}_{\textit{13}}\bra{\Phi^+}+(1-\nu_\pi)\frac{\Id_{\textit{13}}}{4})\otimes \ket{\Psi^+}_{\textit{24}}\bra{\Psi^+},
\end{equation}
where $\nu_\pi$ is the visibility in the polarization degree of freedom, and $\ket{\Psi^+}_{\textit{24}}$ represents the noiseless initial state in the path degree of freedom. The polarization visibility can be measured experimentally leading to $\nu_\pi=0.858\pm 0.008$.

\begin{figure}[h]
\includegraphics{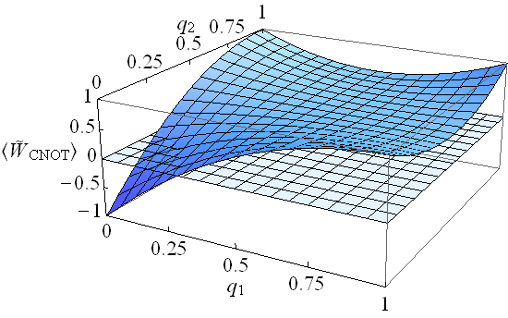}
\caption{(color online) Theoretical expectation value $\langle\tilde W_\cnot\rangle$ as a function of the two noise parameters $q_1$ and $q_2$ for a pure initial Choi state. The plane $\langle\tilde W_\cnot\rangle =0$ is plotted for a comparison.}
\label{dephasing}
\end{figure}

The second source of noise instead affects the path degree of freedom just before the measurement and models the non perfect superposition of modes at the BS level. The net experimental effect is a non perfect interference that can be theoretically described by  the following dephasing channel $\map{D}_{\text{BS}}$:
\begin{align}
\rho\rightarrow (1-\eta_k)^2\rho
	+\eta_k(1-\eta_k)&[\sigma_{z}^{(\textit{2})}\rho\sigma_{z}^{(\textit{2})}
		+\sigma_{z}^{(\textit{4})}\rho\sigma_{z}^{(\textit{4})}]\\
&				+\eta_k^2\sigma_z^{(\textit{2})}\sigma_z^{(\textit{4})}\rho
				\sigma_z^{(\textit{2})}\sigma_z^{(\textit{4})},\nonumber
\end{align}
where the parameter $\eta_k$ represents the strength of the dephasing process and can be connected to the path visibility $\nu_k$ via the formula $\nu_k=(1-2\eta_k)^2$ \cite{105}. A path visibility $\nu_k=0.904\pm0.004$ is measured, hence an experimental value of about $\eta_k=0.025$ is found.

\begin{figure*}[t!]
\includegraphics{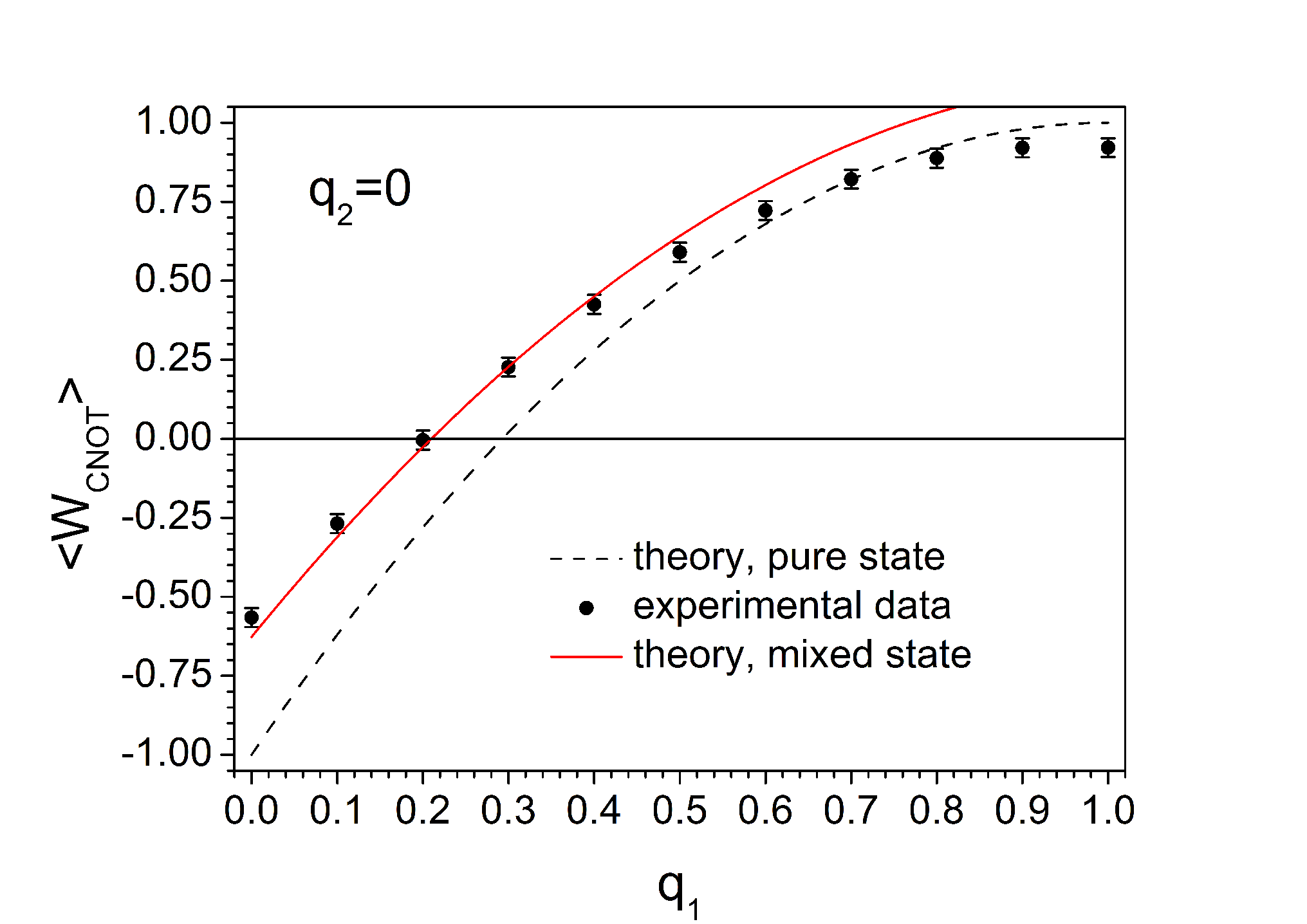}\includegraphics{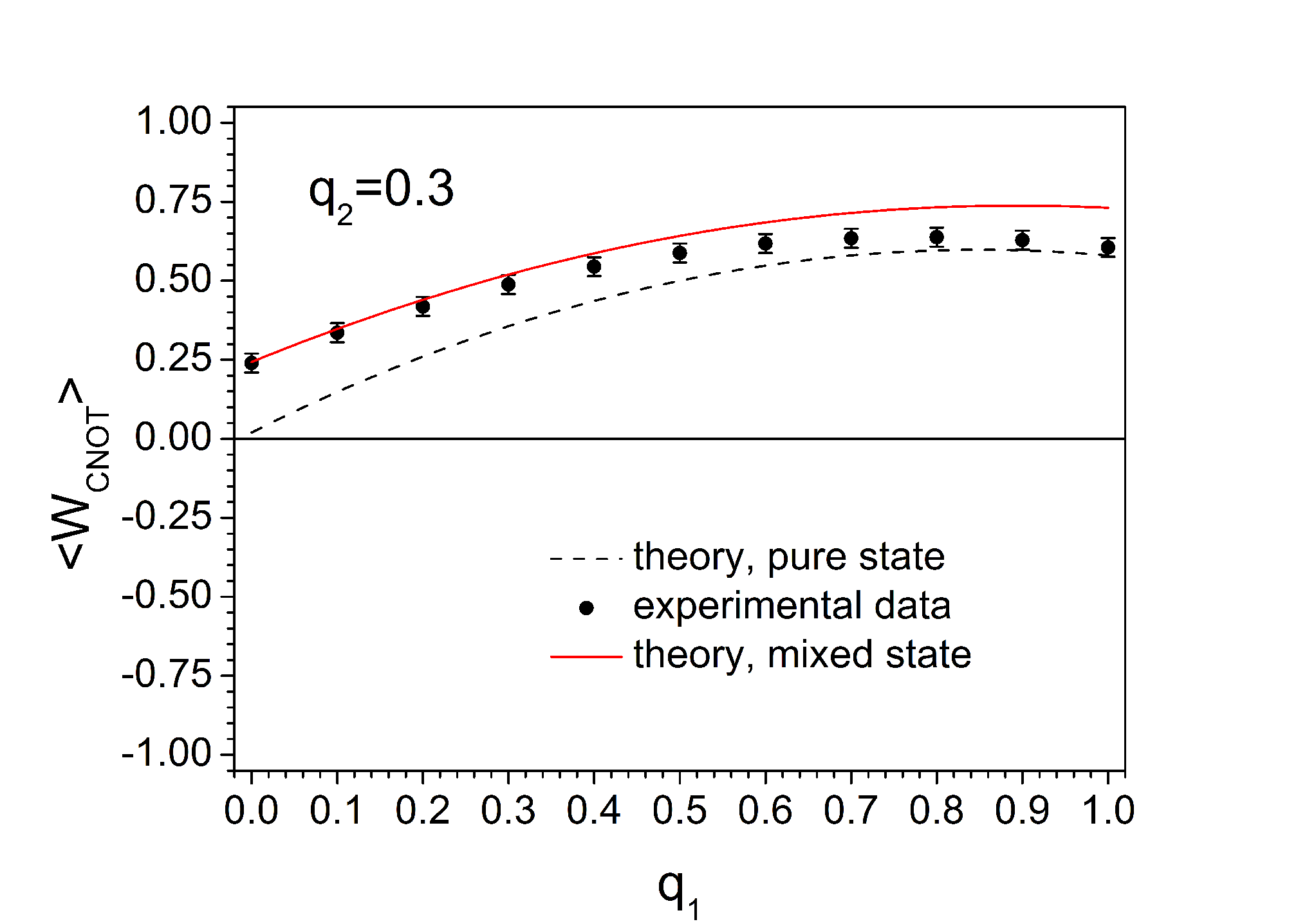}
\caption{(color online) CNOT witness value as a function of the noise parameter $q_1$, in two cases: $q_2=0$ and $q_2=0.30$.}
\label{Wcnotq1q2}
\end{figure*}

Therefore, instead of the perfect Choi state $C_{\map{M}_{\cnot, \map{D}}}$ (see main text for details), we experimentally implemented the following noisy four-qubit state
\begin{equation}\label{noisy}
\rho_{out}=\map{D}_{\text{BS}}\circ\map{M}_{\cnot, \map{D}}[\rho_{in}],
\end{equation}
with respect to which we calculate the expectation value of the witness $\tilde W_\cnot$ as a function of $q_1$, $q_2$, $\nu_\pi$ and $\eta_k$. We recall the reader that the two parameters $q_1$ and $q_2$ are controlled experimentally and can be changed in order to make the $\cnot$ gate nosier, while $\nu_\pi$ and $\eta_k$ model the unavoidable noise present in the experimental set-up. In the case with $q_1=q_2=q$, i.e. when the controlled dephasing processes before and after the CNOT gate have the same strength, the following expectation value of $\tilde W_\cnot$ over the state $\rho_{out}$ given by Eq. \eqref{noisy} can be found:
\begin{align}
\Tr[\tilde W_\cnot \rho_{out}]&= 1 - 2\nu_\pi \\
&+2[2q^3+\eta_k(\eta_k-1)(2q-1)^3](1+\nu_\pi)\nonumber \\
&-2q^2(3+4\nu_\pi)+q(3+5\nu_\pi) \nonumber.
\end{align}
By setting the measured values of $\nu_\pi$ and $\eta_k$, the above expression is exploited to fit the experimental data of $\tilde W_\cnot$ in terms of $q$ (see main text for further details).

\textit{CNOT witness for $q_1 \neq q_2$ ---}
The theoretical expectation value of $\langle\tilde W_\cnot\rangle$ is plotted in Fig. \ref{dephasing} as a function of the dephasing noise parameters $q_1$ and $q_2$ for a pure initial Choi state. As can be seen, the detection on the non-separability of the map can be achieved for sufficiently low noise levels. In particular, if one of the channels is noiseless (e.g. $q_2=0$), the map is non separable for $0\leqslant q_1 <0.29$ \cite{ns1}.

The experimental results for the case $q_1=q_2$ have already been presented in the main text of this letter; in Fig.\ref{Wcnotq1q2}, we present two additional measurements of $\langle\tilde W_\cnot\rangle$ for which only $q_1$ was varied, in the cases $q_2=0$ and $q_2=0.30$. The expectation value for the imperfect initial Choi state prepared was calculated using the model described above. Here again we believe the discrepancy with the experimental data for large $q_1$ may be due to imperfections in the simulated dephasing noise. In the case $q_2=0$, the non-separability of the map is no longer detected for $q_1\geqslant 0.21$ in our experiement; while for the case $q_2=0.30$, as expected (as $q_2>0.29$), the entanglement of the CNOT gate is never detected, whatever the strength $q_1$ of the noise before the gate.